\documentclass[aps,reprint,superscriptaddress,citeautoscript,floatfix,longbibliography,bibnotes]{revtex4-1}
\usepackage{times,siunitx,revquantum,xfrac}
\usepackage[utf8]{inputenc}
\usepackage[T1]{fontenc}
\usepackage[capitalize]{cleveref}
\newcommand{\orcid}[1]{\href{https://orcid.org/#1}{\hspace{1pt}\includegraphics[width=8pt]{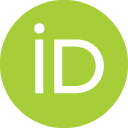}}\, }

\begin{document}
\title{Topologically quantized current in quasiperiodic Thouless pumps}
\author{Pasquale Marra\orcid{0000-0002-9545-3314}}
\email{pmarra@ms.u-tokyo.ac.jp}
\affiliation{
Graduate School of Mathematical Sciences,
The University of Tokyo, 3-8-1 Komaba, Meguro, Tokyo, 153-8914, Japan}
\affiliation{
Department of Physics, and Research and Education Center for Natural Sciences, 
Keio University, 4-1-1 Hiyoshi, Yokohama, Kanagawa, 223-8521, Japan}
\author{Muneto Nitta\orcid{0000-0002-3851-9305}}
\affiliation{
Department of Physics, and Research and Education Center for Natural Sciences, 
Keio University, 4-1-1 Hiyoshi, Yokohama, Kanagawa, 223-8521, Japan}
\date{\today}

\begin{abstract}
Thouless pumps are topologically nontrivial states of matter with quantized charge transport, which can be realized in atomic gases loaded into an optical lattice.
This topological state is analogous to the quantum Hall state.
However, contrarily to the exact, extremely precise, and robust quantization of the Hall conductance, the pumped charge is strictly quantized only when the pumping time is a multiple of a characteristic timescale, i.e., the pumping cycle duration.
Here, we show instead that the pumped \emph{current} becomes exactly quantized, independently from the pumping time, if the system is led into a quasiperiodic, incommensurate regime.
In this quasiperiodic and topologically nontrivial state, the Bloch bands and the Berry curvature become flat, the pumped charge becomes linear in time, while the current becomes steady, topologically quantized, and proportional to the Chern number.
The quantization of the current is exact up to exponentially small corrections.
This has to be contrasted with the case of the commensurate (nonquasiperiodic) regime, where the current is not constant, and the pumped charge is quantized only at integer multiples of the pumping cycle.
\end{abstract}
\maketitle


The hallmark of topological states of matter is the exact quantization of a physical observable in terms of a conserved quantity, the topological invariant~\cite{thouless_quantized_1982,thouless_quantization_1983}.
A paradigmatic example is the quantum Hall conductance, which is quantized as integer (or fractional) multiples of $e^2/h$ with a precision exceeding one part in a billion~\cite{klitzing_new_1980,laughlin_quantized_1981}.
Moreover, this quantization is robust against perturbations, i.e., it persists in the presence of disorder, defects, impurities, or imperfections of the experimental sample.
This led to an extremely precise definition of the electrical resistance standard and experimental determination of the finite-structure constant~\cite{klitzing_quantum_2017}.

A topologically equivalent state is the Thouless pump~\cite{thouless_quantization_1983,niu_towards_1990,shindou_quantum_2005,fu_time_2006,wei_anomalous_2015,roux_quasiperiodic_2008,wang_topological_2013,marra_fractional_2015,marra_fractional_2017,matsuda_two-dimensional_2019}, which can be engineered, e.g., with ultracold atoms~\cite{lewenstein_ultracold_2007,bloch_many-body_2008,zhang_topological_2018,cooper_topological_2019} in a superlattice created by the superposition of two optical lattices with different wavelengths~\cite{nakajima_topological_2016,lohse_thouless_2016,taddia_topological_2017,das_realizing_2019}.
When the superlattice is adiabatically and periodically varied in time $t$, the charge pumped through the atomic cloud is quantized in terms of the topological invariant, i.e., the Chern number~\cite{thouless_quantization_1983}.
However, the charge is quantized only when the duration of the pumping process is an integer multiple of the full adiabatic cycle, and deviations from the quantized value are linear in time. 
In this sense, the quantization of the pumped charge is not exact:
This constitutes a fundamental hindrance to the realization of metrological standards.

\begin{figure}[t]
\centering
\includegraphics[width=1\columnwidth]{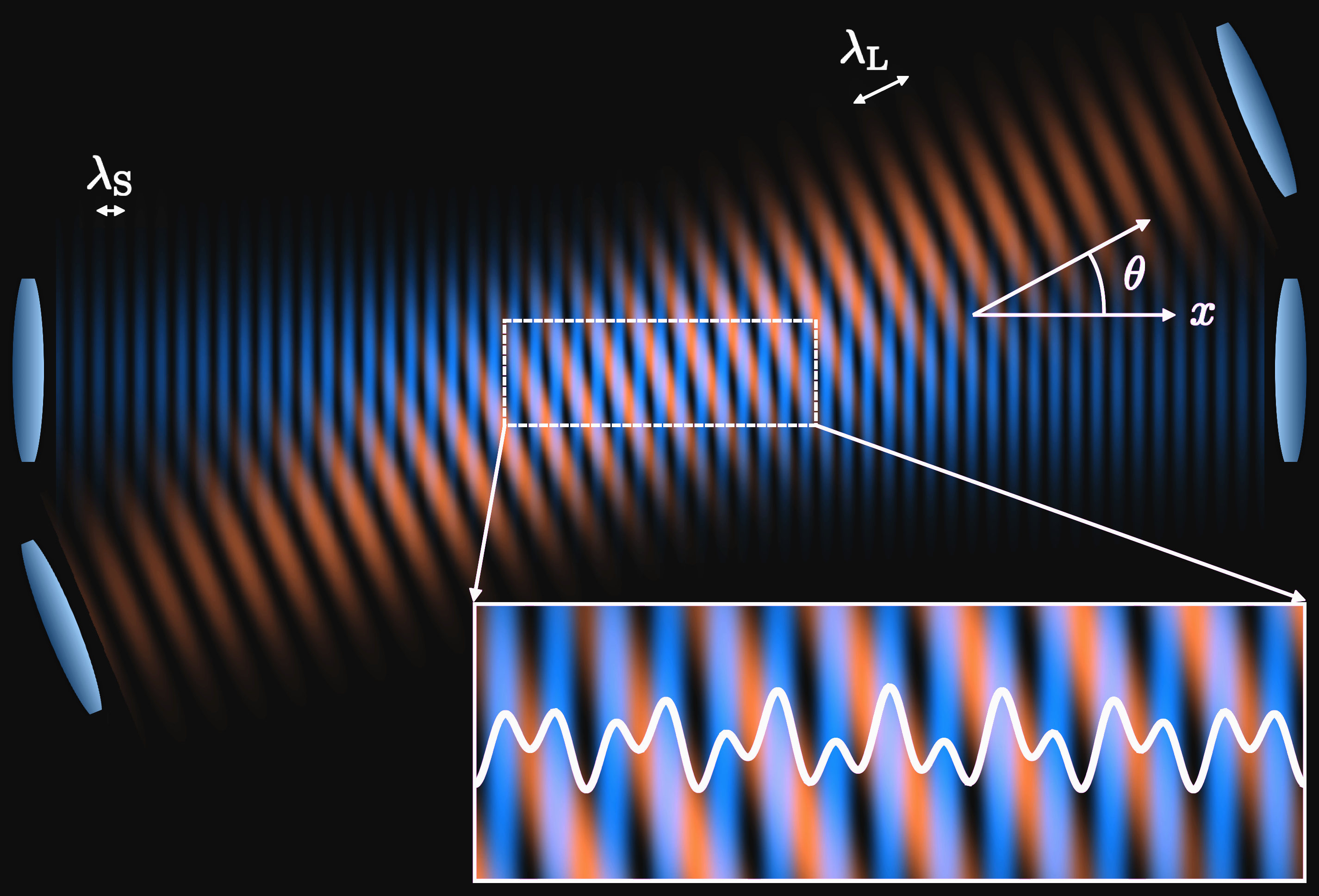}%
\setlength{\belowcaptionskip}{-14pt}\setlength{\abovecaptionskip}{2pt}
	\caption{%
The superposition of two stationary lattices in a tilted direction produces a quasiperiodic one-dimensional lattice when $\alpha=\lambda_\mathrm{S}/(\lambda_{\mathrm{L}}\cos{\theta})$ is an irrational number.
}
	\label{fig1}
\end{figure}

In this Rapid Communication, we will show that the quantization of the pumped \emph{current} can be indeed realized by Thouless pumps in the \emph{quasiperiodic} regime and, most importantly, that this quantization is exact.
In ultracold atomic systems, quasiperiodicity~\cite{kraus_topological_2012-1,kraus_topological_2012-2,kraus_quasiperiodicity_2016,ozawa_topological_2019,valiente_super_2019,kuno_disorder-induced_2019,yao_critical_2019} is realized using a superposition of two optical lattices with incommensurate lattice constants, i.e., their ratio $\alpha$ is an irrational number.
In this regime, the translational symmetry is completely broken, the familiar concept of Brillouin zone (BZ) becomes ill-defined, and the usual definition of the Chern number as an integral of the Berry curvature breaks down.
In order to consider a realistic experimental setup, we will derive an effective tight-binding (TB) model describing an atomic gas in a bichromatic potential~\cite{das_realizing_2019,roux_quasiperiodic_2008}, which coincides with a generalized Aubry-Andr\'{e}-Harper-Hofstadter (AAHH) model~\cite{harper_single_1955,hofstadter_energy_1976,aubry_analyticity_1980,hatsugai_energy_1990,osadchy_hofstadter_2001,hatsuda_hofstadters_2016,ikeda_hofstadters_2018} with an extra spatially dependent tunneling term.
Furthermore, we will operatively define the Chern number by taking the limit of an ensemble of periodic and topologically equivalent states which progressively approximate quasiperiodicity. 
In this limit, the Bloch bands and Berry curvatures become asymptotically flat, as already known~\cite{kraus_topological_2012-1,harper_perturbative_2014}.
Finally, we describe the experimental fingerprint of the quasiperiodic topological state, which reveals itself in the charge transport and adiabatic evolution of the center of mass of the atomic cloud.
Whereas in the commensurate (nonquasiperiodic) case the current is not constant and the pumped charge is quantized only at exact multiples of the pumping cycle, we find that the quasiperiodic nontrivial state is characterized by a steady and topologically quantized pumping current, independently from the duration of the pumping process.
Most importantly, we find that this quantization is exact up to exponentially small corrections, it is robust against perturbations which do not break the symmetries of the system, and does not depend on the details of the model considered.
This exact quantization is a direct consequence of quasiperiodicity, and may contribute to a more accurate definition of current standards~\cite{kaneko_review_2016}.


\begin{figure}[t]
\centering
\includegraphics[width=1\columnwidth]{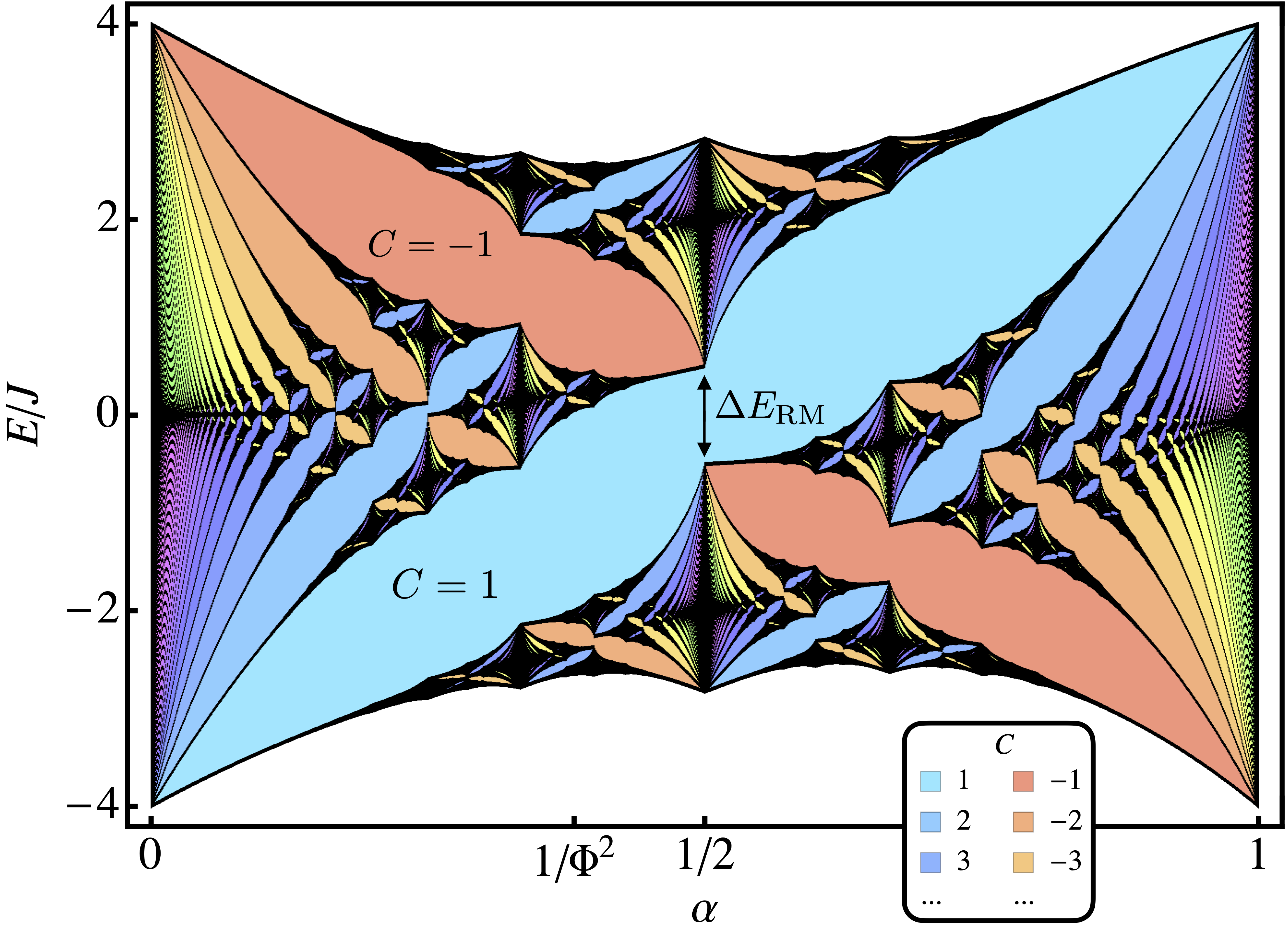}%
\setlength{\belowcaptionskip}{-4pt}\setlength{\abovecaptionskip}{2pt}
	\caption{%
Energy spectra of the TB Hamiltonian \eqref{Hk} calculated for $V=J$ and $K=0.25J$. 
The large central gap has Chern number $C=1$ and is topologically equivalent to the RM model ($\alpha=1/2$).
For $K\to0$ (not shown) the central gaps close at $\alpha=p/q$ for $q$ even.
Different color shades correspond to different Chern numbers.
}
	\label{fig2}
\end{figure}

Experimentally, Thouless pumps are realized by ultracold Fermi gases loaded into dynamically controlled bichromatic lattices~\cite{nakajima_topological_2016,lohse_thouless_2016}.
Using a tilted setup~\cite{nakajima_disorder_2020,matsuda_two-dimensional_2019} as in \cref{fig1}, two sets of counterpropagating laser beams produce two standing waves with wavelengths $\lambda_\mathrm{S}$ and $\lambda_\mathrm{L}>\lambda_\mathrm{S}$ which intersect at an angle $\theta$.
For an atomic cloud confined in the $x$ direction, the total dipole potential is
\begin{equation}
V(x,\phi)= 
 V_\mathrm{S}\cos^2\left(\frac{\pi x}{d_\mathrm{S}}\right)
+ 
V_\mathrm{L}\cos^2\left(\frac{\pi x}{d_\mathrm{L}}-\frac\phi2 \right),
\label{CH}
\end{equation}
where 
$d_\mathrm{S}=\lambda_\mathrm{S}$ and $d_\mathrm{L}=\lambda_\mathrm{L}\cos{\theta}$ are respectively the short and long lattice constants,
$V_\mathrm{S,L}$ the lattice depths, and $\phi$ the phase difference between the two lattices, which varies 
in time as $\phi=\nu t$ with instantaneous frequency $\nu$.
The commensuration $\alpha=d_\mathrm{S}/d_\mathrm{L}=\lambda_\mathrm{S}/(\lambda_\mathrm{L}\cos{\theta})$ between the two lattices is controlled by the tilting angle $\theta$.
We assume a deep lattice regime $V_\mathrm{S}> E_r$ (here, $E_r=h^2/(8 M d_\mathrm{S}^2)$ is the recoil energy of the short lattice~\cite{bloch_many-body_2008}).
If $V_\mathrm{S}> V_\mathrm{L}$, the continuum Hamiltonian ${\cal H}=p^2/2M +V(x,\phi)$ can be discretized using localized states at the short lattice minima and treating the long lattice as a perturbation~\cite{roux_quasiperiodic_2008}.
This leads to an effective low-energy TB Hamiltonian corresponding to a generalized Harper equation which reads
\begin{align}
&
[-J 
\!+\! 
2K \alpha\sin{(\pi\alpha)} \cos(2\pi\alpha (n +1) \!-\! \phi) ] (\psi_{n-1} \!+\! \psi_{n+1})
\,+\nonumber\\
&\qquad
+ 2V\cos(2\pi\alpha (n+1/2)- \phi) \psi_{n} = E \psi_{n}.
\label{HHe}
\end{align}
This is a generalization of the AAHH model, which includes an extra site-dependent tunneling term $K\propto V_\mathrm{L}$.
Moreover, for $\alpha=1/2$ (staggered field), \cref{HHe} reduces to the Rice-Mele (RM) model~\cite{rice_elementary_1982,rice_mele_shen_topological_2017}
\begin{align}
&
 [-J \!-\! K (-1)^n \cos\phi ] (\psi_{n-1} \!+\! \psi_{n+1})
\nonumber\\
&\qquad
+ 2V(-1)^n\sin\phi \,\psi_{n} = E \psi_{n},
\label{eq:RM}
\end{align}
which has an energy gap $\Delta E_\mathrm{RM}=4\min(|J|,|V|,|K|)$.

In the commensurate case, i.e., $\alpha=p/q$ with $p, q$ integer coprimes, one can verify that \cref{CH,HHe} are invariant up to translations $n\to n+q$, and consequently the superlattice unit cell has length $q d_\mathrm{S}$.
In momentum space,
\begin{align}
&\qquad
H= 
\sum_k -2J\cos{k}\, c_k^\dag c_k
+
 e^{\ii (\pi \alpha-\phi)}
 \nonumber\\\times
 &
 \left[
 V \!+\! 2K \alpha\sin{(\pi\alpha)}
\cos{(k\!+\!\pi\alpha)}
\right]\!
c^\dag_{k} c_{k+2\pi\alpha}
\!+\! \text{H.~c.},
\label{Hk}
\end{align}
where $k$ is restricted to the first BZ $[0,2\pi/q]$.
\Cref{fig2} shows the energy spectra of the TB model, which are a deformed version of the Hofstadter butterfly~\cite{hofstadter_energy_1976,avila_ten_2009}.
Indeed, whereas the Hofstadter butterfly ($K=0$) is symmetric with respect to the transformations $\alpha\to1-\alpha$ and $E\to -E$ (corresponding to $k\to k+\pi$), the spatially dependent tunneling term breaks these symmetries.
For small $K$, one can assume that the intraband gaps remain open for $K\to0$ and are thus homeomorphic to the gaps of the Hofstadter butterfly.
Thus, the intraband gaps are topologically nontrivial with Chern number $C\neq0$ satisfying the diophantine equation $p C \equiv j \mod q$ (analogously to the Hofstadter butterfly $K=0$).
Unlike the original Hofstadter butterfly, the energy spectra is gapped at $E=0$ for $\alpha=p/q$ with $q$ even.
Intraband gaps with low Chern numbers are generally wide and remain open for a broad range of the commensuration $\alpha$.
In particular, the large central gap in \cref{fig2} is open for any value of $\alpha$ and is topologically equivalent to the RM model: 
It can be continuously deformed into $\alpha\to1/2$, where \cref{HHe} reduces to \cref{eq:RM}.


In the commensurate case, assuming homogeneously populated bands below the Fermi level $E_\mathrm{F}$ and at zero temperature, the total charge pumped during an adiabatic evolution $\phi\to \phi+ 2\pi $ is quantized and equal to the Chern number $C$ of the filled Bloch bands~\cite{thouless_quantization_1983}
$
Q=C=
(1/2\pi)
\int_{\phi}^{\phi+2\pi} \dd \phi
\int_\mathrm{0}^{2\pi/q} \dd k
\Omega
$.
Here, 
$\Omega
=\sum_i
\Theta(E_\mathrm{F}-E_i)
\omega_i
$ 
is the total Berry curvature at the Fermi level $E_\mathrm{F}$, 
with
$\Theta(E)$ the Heaviside step function and 
$\omega_i
=2\Im
\braket{\partial_\phi u_i | \partial_k u_i}$
the Berry curvature of the $i$-th band, defined in terms of the Bloch wavefunctions 
$\ket{\psi_{i}(k,x)}=e^{\ii k x}\ket{u_i(k,x)}$.
Moreover, the current 
$
I=\partial_\phi Q=(1/2\pi)
\int_\mathrm{0}^{2\pi/q} \dd k
\Omega
$
is not quantized and not constant during the pumping process, oscillating around an average value 
$\langle I \rangle=
\langle\Omega
\rangle/{q}$ 
with maximum variation
$\delta I \leq 
\delta\Omega
/{q}$ 
where $\delta\Omega
=\max \Omega
-\min \Omega
$.

Due to translational invariance, Hamiltonian \eqref{Hk} is periodic in the momentum $k\to k+2\pi/q$, but not in the phase since $H(\phi+2\pi/q)\neq H(\phi)$.
One can show that a phase shift $\phi\to\phi+2\pi m/q$ in \cref{CH,HHe} is equivalent to a translation $n\to n-c$, where $c$ satisfies the diophantine equation $p c\equiv m\mod q$.
Thus, the Hamiltonian is ``unitarily'' periodic~\cite{marra_fractional_2015,marra_fractional_2017} in the phase $\phi$ up to lattice translations, i.e., it is periodic up to unitary transformations (translations),
\begin{equation}
\label{translation}
 H(\phi+2\pi m/q)= {T}^{-c} H(\phi) {T}^{c}, 
\end{equation}
where ${T}$ is defined by $T V(x,\phi)T^{-1}= V(x+d_\mathrm{S},\phi)$.
Consequently, energies and Berry curvatures are periodic in the phase $\phi\to\phi+2\pi/q$, and the pumped charge at well-defined fractions of the pumping cycle $\Delta\phi=2\pi m/q$ is quantized as fractions of the Chern number~\cite{marra_fractional_2015,marra_fractional_2017}
$
Q=m C/q=
(1/2\pi)
\int_{\phi}^{\phi+2\pi m/q} \dd \phi
\int_\mathrm{0}^{2\pi/q} \dd k\,
\Omega
$.
Moreover, the energy bands, Berry curvatures, and  total Berry curvature become flat in the limit of large denominators $q$.


\begin{figure}[t]
\centering
\includegraphics[width=1\columnwidth]{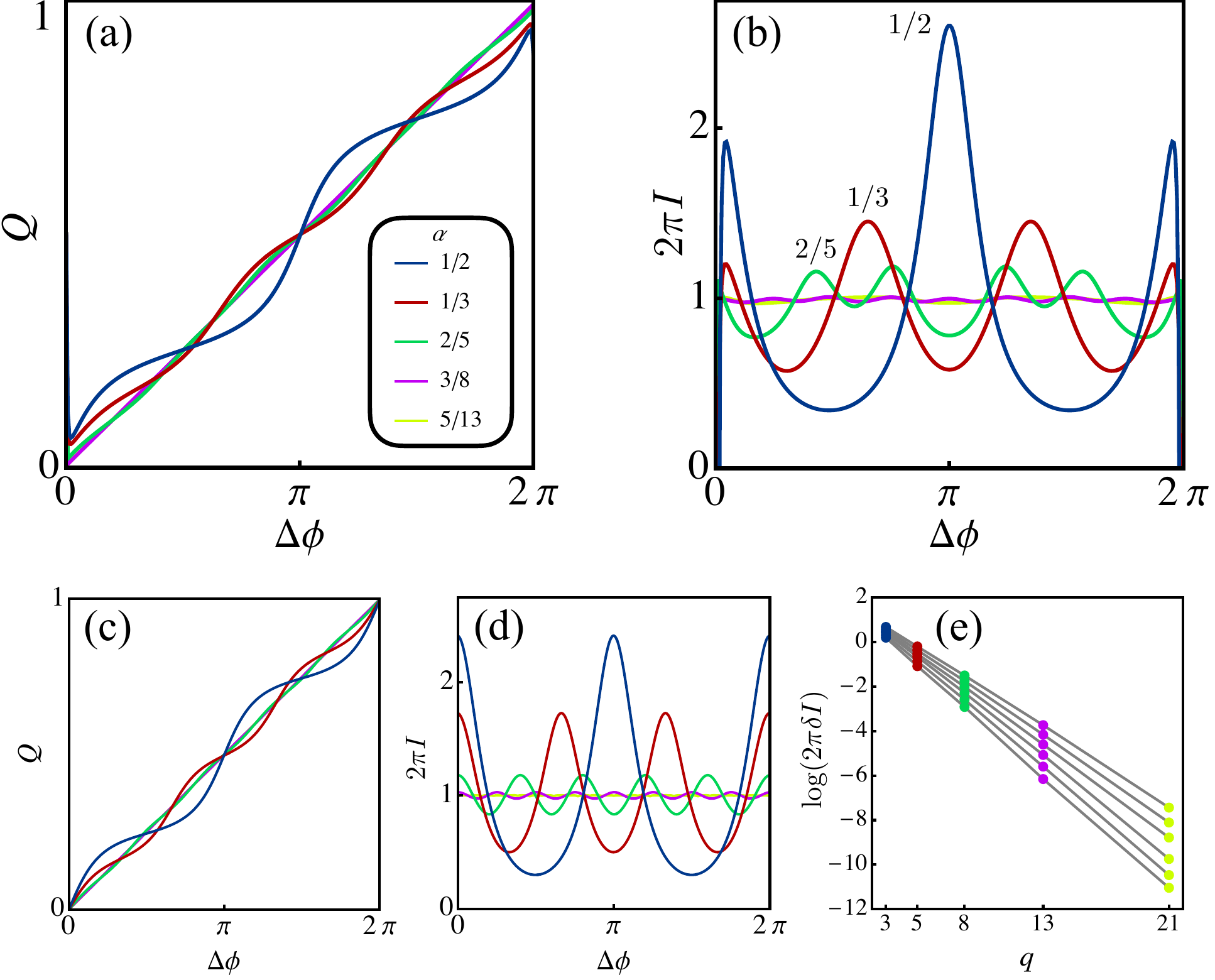}%
\setlength{\belowcaptionskip}{-4pt}\setlength{\abovecaptionskip}{2pt}
	\caption{%
Pumped charge and current for the central gap with Chern number $C=1$ calculated with the continuous model (a,b) and the effective TB model (c,d) respectively.
Different curves correspond to successive rational approximations of $\alpha=1/\Phi^2$, corresponding to tilting angles $\theta=\arccos{(1/(4\alpha))}$ in the range between $\ang{60}$ and $\ang{49}$ for typical laser wavelengths $\lambda_\mathrm{S}=\SI{266}{nm}$ and $\lambda_\mathrm{L}=\SI{1064}{nm}$.
The pumped charge is quantized as $m C/q$ for pumping periods $\Delta\phi=2\pi m/q$.
In the limit $\alpha_n\to\alpha$, the charge has a linear dependence $Q=\Delta \phi C_\alpha/2\pi$.
The current shows large fluctuations but becomes steady for large denominators $q$, reaching its quantized value $I_\alpha=C/2\pi$.
We use $V_\mathrm{S}=2 E_r$, $V_\mathrm{L}=0.5 E_r$.
(e)
The current approaches its quantized value exponentially as $\delta I=|I-I_\alpha|\propto\exp(-q/\xi)$.
Different data sets correspond to $V=J, 1.25 J$ and $K=0.25 J, 0.375 J, 0.5 J$.
}
	\label{fig3}
\end{figure}

We now consider the incommensurate quasiperiodic case $\alpha\in\mathbb R-\mathbb Q$.
Every irrational number $\alpha$ can be written uniquely as an infinite continued fraction~\cite{continued_fractions_hardy} 
$\alpha=[a_0; a_1, a_2, \,\ldots ] =
a_0+
1/(a_1 + 
1/(a_2 + 
\dots))$
with $a_i$ integers.
Successive approximations obtained by truncating the continued fraction representation $\alpha_n=[a_0; a_1, a_2, \,\ldots, a_n]$ are rational numbers, and converge to $\alpha$.
We thus consider the ensemble of Hamiltonians $H^{(\alpha_n)}$ describing commensurate systems with $\alpha_n= p_n/q_n=[a_0; a_1, a_2, \,\ldots, a_n]$.
We assume that the insulating gap at the Fermi level remains open, such that the Hamiltonians $H^{(\alpha_n)}$ are topologically equivalent.
As the denominator $q_n$ increases for $n\to\infty$, the BZ $[0,2\pi/q_n]$ shrinks and becomes ill-defined in the quasiperiodic limit.
Thus, the usual definition of the Chern number as an integral of the Berry curvature in the BZ needs to be reformulated.
However, energy bands and Berry curvatures become constant in the quasiperiodic limit ($q_n\to\infty$).
Hence, if the gap remains open for $\alpha_n\to\alpha$, the Berry integral converges for $n\to\infty$, and we can define the Chern number in the quasiperiodic limit as
\begin{equation}
C_\alpha=\!\!\lim_{\alpha_n\to\alpha}
\frac1{2\pi} \!
\int_{0}^{2\pi} \!\!\!\! \!\!\! \dd \phi \!
\int_\mathrm{0}^{{2\pi}/{q_n}} \!\!\! \!\!\! \dd k\,
\Omega^{(\alpha_n)} 
=\!\!
\lim_{\alpha_n\to\alpha}
\frac{2\pi}{q_n}\, \Omega^{(\alpha_n)} 
.
\label{chern}
\end{equation}
In this limit, the Chern number is simply proportional to the total Berry curvature, which diverges asymptotically as $\Omega^{(\alpha_n)}\sim q C/2\pi$.
Moreover, since the total Berry curvature is flat, the charge pumped during adiabatic transformations, for any initial and final values of the phase $\phi\to\phi+\Delta\phi$, becomes
\begin{equation}
\label{charge}
Q_\alpha=
\!\!\lim_{\alpha_n\to\alpha}
\frac1{2\pi} \!
\int_{\phi}^{\phi+\Delta\phi} \!\!\!\! \!\!\! \dd \phi \!
\int_\mathrm{0}^{{2\pi}/{q_n}} \!\! \dd k\,
\Omega^{(\alpha_n)} 
=
\frac{\Delta\phi}{2\pi} C_\alpha,
\end{equation}
whereas the instantaneous charge current becomes
\begin{equation}
\label{current}
I_\alpha=
\!\!\lim_{\alpha_n\to\alpha}
\frac1{2\pi}
\int_\mathrm{0}^{{2\pi}/{q_n}} \!\! \dd k\,
\Omega^{(\alpha_n)} 
=
\frac{C}{2\pi}
\end{equation}
In the quasiperiodic limit, the pumped charge becomes linear in the phase difference $\Delta\phi$, whereas the current $I=\partial_\phi Q$ becomes constant and proportional to the Chern number. 
Notice that, in order to observe the effects of quasiperiodicity, the system size $L$ must be larger than the unit cell $q d_\mathrm{S}$.
In this sense, the limit $\alpha_n\to\alpha$ corresponds to the infinite-size limit $L\to\infty$.

These effects are robust against perturbations which do not break translational symmetry.
In fact, adding a perturbation $\lambda V$ in \cref{translation}, one can verify that the perturbed Hamiltonian satisfies
\begin{equation}
\label{brokentranslation}
 H'(\phi+2\pi m/q)= {T}^{-c} (H'(\phi)+c \lambda [T,V] T^{-1}) {T}^{c}.
\end{equation}
If translational symmetry is unbroken, this equation reduces to \cref{translation}.
In this case, energy levels and Berry curvatures are still periodic and become flat in the quasiperiodic limit, and the current remains quantized.
However, if $[V,T]\neq0$, from \cref{brokentranslation} one can expect polynomial corrections $O(\lambda)$ to the energy levels and Berry curvatures.
Thus, spatial disorder is expected to break down the exact quantization of the current.
However, disorder is usually negligible in optical lattices, contrarily to solid state systems.


We now consider the continuous Hamiltonian $
{\cal H}=
{p^2}/{2M}
+V(x,t)
+ V_\mathrm{T} x^2
$
describing an ultracold atomic cloud in a bichromatic potential, confined by a shallow harmonic trap $\propto V_\mathrm{T}$.
The pumped current $I=\partial_\phi Q=\partial_t Q/\nu$ is related to a simple physical observable, i.e., the center of mass of the atomic cloud.
The variation of the center of mass $\langle x (t)\rangle=(1/N) \sum_{i=1}^j \int_{-\infty}^\infty |\Psi_i(x,t)|^2 x \dd x$ is proportional to the pumped charge~\cite{marra_fractional_2015,wang_topological_2013}, i.e.,
$Q=\rho [\langle x (t+\Delta t)\rangle - \langle x (t)\rangle]$,
where $\rho=j/(q d_\mathrm{S})$ is the number of atoms $j$ per unit cell.
Assuming the number of filled bands to be $j \equiv p C \mod{q}$, the total length of a cloud of $N$ atoms is given by $N/j$ unit cells (of length $q d_\mathrm{S}$).
Hence the number of atoms $N$ must be multiple of the filling factor $j$, and the system length $L$ must be tuned such that
\begin{equation}
\label{condition}
d_\mathrm{S}\frac{N}{L} \equiv \alpha C\mod q.
\end{equation}
Moreover, in order to minimize thermal and nonadiabatic effects, one should consider a filling factor $j=p$  corresponding to the large central gap in \cref{fig2} with Chern number $C=1$.
This gap $\Delta E$ has the same order of magnitude for a wide range of values of the commensuration $\alpha$, including $\alpha=1/2$ where the system is equivalent to the RM model, i.e., $\Delta E \approx \Delta E_\mathrm{RM}$.
This fixes the temperature and timescales to $T<\Delta E_\mathrm{RM}/k_\mathrm{B}$ and $\nu<\Delta E_\mathrm{RM}/\hbar$.
Note that the RM quantum pump has been already realized experimentally~\cite{nakajima_topological_2016,lohse_thouless_2016}. 
Note also that the experimental errors in measuring the center of mass can be reduced by averaging over a large number of cycles~\cite{nakajima_topological_2016,lohse_thouless_2016}.


Figure~\ref{fig3} shows the pumped charge $Q$ and the current $I=\partial_\phi Q$ obtained by calculating the center of mass of the continuous system in the adiabatic limit and, alternatively, using the effective TB Hamiltonian \eqref{Hk}.
Different curves correspond to successive rational approximations of $\alpha=1/\Phi^2\in \mathbb{R}-\mathbb{Q}$, where $\Phi$ is the golden ratio.
We tune the trapping potential such that the length $L$ satisfies \cref{condition}.
The pumped charge is quantized as integer fractions of the Chern number $(m/q) C$ for well-defined fractions of the pumping period $\Delta\phi=2\pi m/q$.
For increasing denominators $q$, the pumped charge is approximately $Q=\Delta \phi C_\alpha/2\pi$, whereas the current approaches its quantized value $I_\alpha=C_\alpha/2\pi$ for $\alpha_n\to\alpha$.


Hence, the pumped current $I_\alpha$ in the quasiperiodic limit is quantized and equal to the Chern number (in elementary units).
We will now determine the asymptotic behavior of the current approaching the quasiperiodic limit.
For $K=0$, \cref{Hk} reduces to the AAHH model: 
In this case, it has been shown numerically and perturbatively~\cite{harper_perturbative_2014} that the total Berry curvature takes the form 
$
\Omega^{(p/q)}\approx F+G e^{- q/\xi} [\cos{(q k)}+\cos{(q \phi)}]
$.
It is reasonable to extrapolate this result also to $K\neq0$.
\Cref{chern} gives $F=q C/2\pi$, whereas $G\propto q^2$~\cite{harper_perturbative_2014}.
Hence, the flattening of the total Berry curvature is exponential, and  
$\delta\Omega^{(q)}\approx g q^2 e^{- q/\xi}$  asymptotically for large $q$, where $g>0$ is a constant.
Thus, the current approaches its quantized value as
\begin{equation}
\delta I=|I-I_\alpha|\lesssim
g q \
e^{- q_n/\xi}
\approx
\frac{g \ e^{- \sfrac{1}{\xi \sqrt{D|\alpha-\alpha_n|}}}}{\sqrt{D |\alpha-\alpha_n|}},
\label{scaling}
\end{equation}
where $|\alpha-\alpha_n|\sim 1/{D q_n^2}$ with $D<\sqrt{5}$, due to the Dirichlet's approximation theorem and Hurwitz's theorem~\cite{continued_fractions_hardy}.
Thus, \cref{scaling} describes the scaling behavior of the current in the quasiperiodic limit, in terms of the difference $|\alpha-\alpha_n|$ between the irrational commensuration $\alpha$ and its successive rational approximations $\alpha_n=p_n/q_n$.
The denominator $q_n$ determines the length scale $L_n=q_n d_\mathrm{S}$ where the effects of quasiperiodicity become relevant.
Consequently, \cref{scaling} mandates that corrections to the quantized value of the current are exponentially small in the system size $L$.
This is a distinctive fingerprint of topological quantization, and is analogous to the case of, e.g., the quantum Hall effect, where corrections to the quantized conductance are exponentially small in the linear dimensions of the system~\cite{niu_quantum_1987,exponentially_small_topological_thouless_1998}.
\Cref{fig3}(e) shows the variations $\delta I$ calculated numerically via \cref{current} using the effective TB Hamiltonian \eqref{Hk}.
As expected, the current approaches its quantized value $I_\alpha=C_\alpha/2\pi$ exponentially for $\alpha_n\to\alpha$.


In summary, we have shown how a quasiperiodic and topologically nontrivial Thouless pump can be realized by an atomic gas confined in a quasiperiodic optical lattice, which is a superposition of two harmonic potentials with incommensurate periodicities.
This system is characterized by a topological invariant defined as the limit of the Chern numbers of an ensemble of topologically equivalent and periodic Hamiltonians.
The distinctive fingerprint of this quasiperiodic and topologically nontrivial state is the exact quantization of the current, which is a consequence of the flattening of the Bloch bands and of the Berry curvatures.
This exact quantization is measurable in a typical experimental setting of ultracold atomic gases in optical lattices, and may open new perspectives for a more accurate definition of current standards.

\begin{acknowledgments}
P.~M. thanks Yoshihito Kuno, Michael Lohse, Shuta Nakajima, Yoshiro Takahashi, and Nobuyuki Takei for useful discussions. 
The work of P.~M. is supported by the Japan Science and Technology Agency (JST) of the Ministry of Education, Culture, Sports, Science and Technology (MEXT), JST CREST Grant~No.~JPMJCR19T2, by the (MEXT)-Supported Program for the Strategic Research Foundation at Private Universities ``Topological Science'' (Grant No.~S1511006), and by JSPS Grant-in-Aid for Early-Career Scientists (Grant No.~20K14375).
The work of M.~N.~is partially supported by the Japan Society for the Promotion of Science (JSPS) Grant-in-Aid for Scientific Research (KAKENHI) Grants No.~16H03984 and No.~18H01217 and by a Grant-in-Aid for Scientific Research on Innovative Areas ``Topological Materials Science'' (KAKENHI Grant No.~15H05855) from MEXT of Japan.
\end{acknowledgments}

\end{document}